\documentclass[12pt]{article}
\usepackage[utf8]{inputenc}
\usepackage{textcomp}
\usepackage{graphicx}
\usepackage{amsmath}
\usepackage{array}
\usepackage{multirow}
\usepackage{booktabs}
\usepackage[hypcap=false]{caption}
\usepackage{xspace}
\usepackage{calc}
\usepackage[hidelinks]{hyperref}

\usepackage{lineno}
\usepackage{etoolbox}
\newcommand*\linenomathpatchAMS[1]{%
  \expandafter\pretocmd\csname #1\endcsname {\linenomathAMS}{}{}%
  \expandafter\pretocmd\csname #1*\endcsname{\linenomathAMS}{}{}%
  \expandafter\apptocmd\csname end#1\endcsname {\endlinenomath}{}{}%
  \expandafter\apptocmd\csname end#1*\endcsname{\endlinenomath}{}{}%
}
\expandafter\ifx\linenomath\linenomathWithnumbers
  \let\linenomathAMS\linenomathWithnumbers
  \patchcmd\linenomathAMS{\advance\postdisplaypenalty\linenopenalty}{}{}{}
\else
  \let\linenomathAMS\linenomathNonumbers
\fi
\linenomathpatchAMS{align}

\DeclareFontEncoding{LGR}{}{}
\DeclareFontSubstitution{LGR}{cmr}{m}{n}
\DeclareTextSymbol{\textgamma}{LGR}{103}
\DeclareTextSymbolDefault{\textgamma}{LGR}

\newcommand{\mybar}[1]{\smash{$\bar{\text{#1}}$}}
\newcommand{\tbar}{\mybar{t}}
\newcommand{\ttbar}{t\tbar}
\newcommand{\tttt}{t\tbar{}t\tbar}
\newcommand{\ttW}{t\tbar W}
\newcommand{\ttZ}{t\tbar Z}

\newcommand{\TeV}{\,TeV\xspace}
\newcommand{\sqrts}[1][13]{$\sqrt{s}=#1$\TeV}

\begin{document}

\begin{titlepage}\pagenumbering{Alph}
{\let\thefootnote\relax
\footnotetext{\hspace*{-18pt}%
Copyright 2023 CERN for the benefit of the ATLAS and CMS collaborations. Reproduction of this article or parts of it is allowed as specified in the CC-BY-4.0 license.}}
\rightline{\begin{tabular}{l}
    CMS-CR-2023/285 \\ 
    December 4, 2023 
\end{tabular}}

\vfill
\begin{center}\Large 
    Four top quark production searches and cross section measurements at the LHC
\end{center}
\vfill
\begin{center}
    \href{mailto:niels.vandenbossche@cern.ch}{\textsc{Niels Van den Bossche}} \\
    \textit{Ghent University, Ghent, Belgium}
\end{center}
\begin{center}
    \textsc{on behalf of the ATLAS \& CMS Collaborations}
\end{center}
\vfill
\begin{quotation} 
    Four top quark production is a rare standard model process that has been observed for the first time 
    in proton-proton collisions at \sqrts at the CERN LHC by both the ATLAS and CMS Collaborations.
    Both observations were made in the same-sign dilepton and multilepton final states of the process
    and are presented in this contribution.
    In addition, another measurement of four top production by the CMS Collaboration
    using the all hadronic, one lepton and opposite-sign dilepton final states is presented.
    This is the first result to take a direct look at the all hadronic final state.
\end{quotation}
\vfill
\begin{quotation}\begin{center}
    PRESENTED AT
\end{center}\bigskip\begin{center}\large
    16\textsuperscript{th} International Workshop on Top Quark Physics\\
    24--29 September, 2023
\end{center}\end{quotation}
\vfill
\end{titlepage}
\def\thefootnote{\fnsymbol{footnote}}
\setcounter{footnote}{0}
\pagenumbering{arabic}

\section{Introduction}
Four top quark production is a rare process with a cross section predicted by the standard model (SM) of $13.4^{+1.0}_{-1.8}$ fb at \sqrts~\cite{fourtop_theory}.
Despite the low cross section, the process can serve as an important probe to new physics,
as new massive bosons can dramatically enhance the production cross section.
In addition, it can serve as a probe of the top-Higgs Yukawa coupling, complementary to other measurements of this SM parameter.

In this contribution, three results by the ATLAS and CMS Collaborations~\cite{ATLAS-Experiment,CMS-Experiment} are presented.
The first is a search for four top production by the CMS Collaboration targetting the zero ($0\ell$), one ($1\ell$) and two opposite-sign lepton ($2\ell$OS) final states~\cite{TOP-21-005}.
Two other results, one from each collaboration, present the observation of this rare process~\cite{ATLAS_FourTop,TOP-22-013}.
These observations were achieved using the two same-sign ($2\ell$SS), three ($3\ell$) and four lepton ($4\ell$) final states.

\section{\texorpdfstring{The $0\ell$, $1\ell$ and $2\ell$OS final states}{The 0l, 1l and 2lOS final states}}
The CMS Collaboration performed a search for four top production in $0\ell$, $1\ell$ and $2\ell$OS events~\cite{TOP-21-005}.
These three final states combined have a large branching fraction of 87\%, but the sensitivity is limited by the background from \ttbar{}
with additional (b) jets and QCD multijet production.
To reduce these contributions, top taggers are used, targetting both boosted and resolved top quark signatures.
The signal and control region definitions differ in each final state, but are generally defined by
a large number of (b tagged) jets and the existence of tagged top quark candidates.

\begin{center}
\centering
\includegraphics[width=0.47\textwidth]{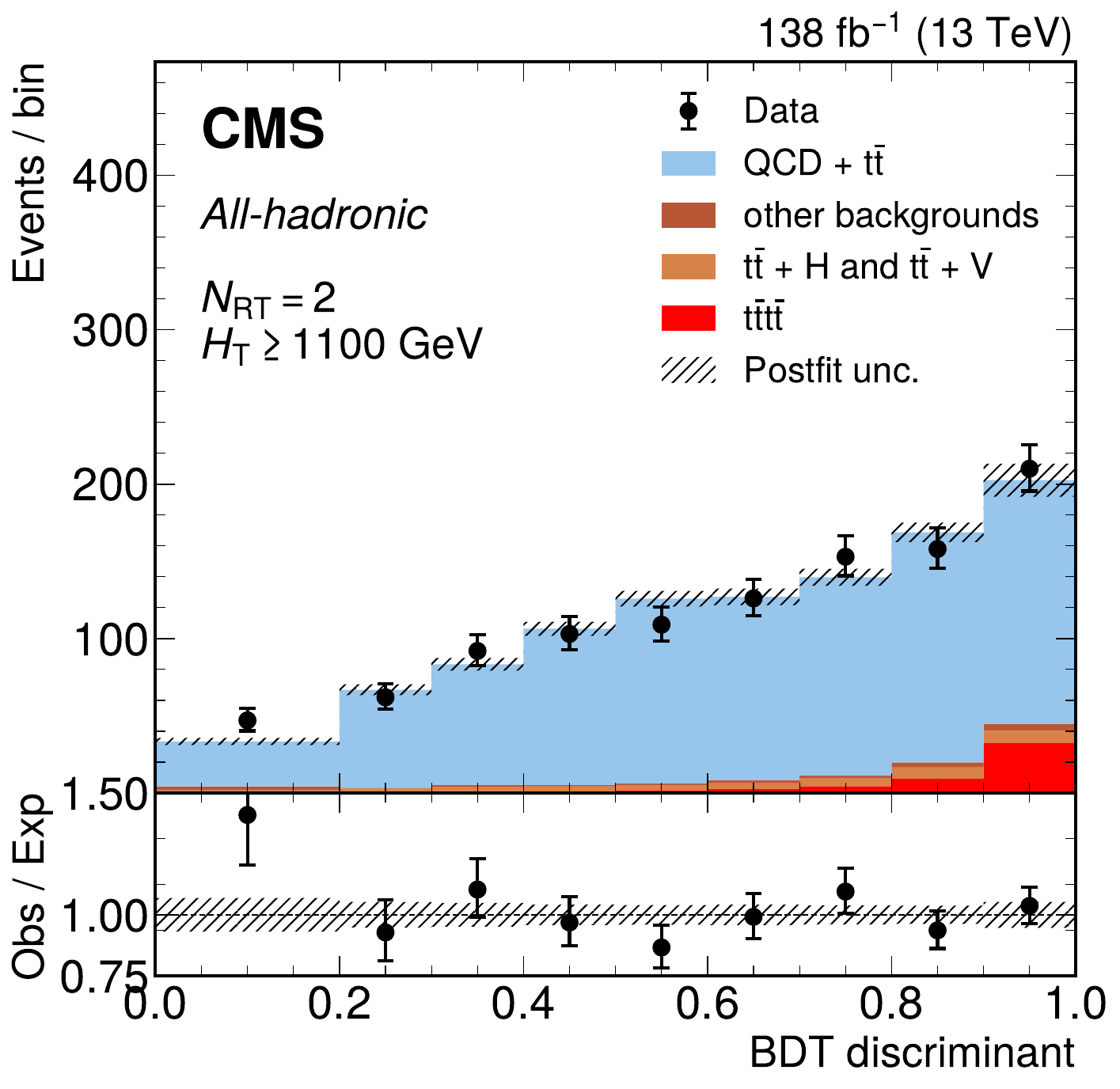}%
\captionof{figure}{%
    The BDT score in the zero lepton (also referred to as all-hadronic) final state. Signal regions like this are used to extract the four top cross section~\cite{TOP-21-005}.
}
\label{fig:cms-summ-allhadro}
\end{center}
Regions enriched in four top events are created using BDTs in the zero and one lepton final states and used in the template fit.
In the two opposite-sign lepton final state, the scalar sum of the transverse momentum of all jets is used as the 
discriminating variable between signal and background.
The cross section is extracted with a template fit to the signal and control regions, with one of the $0\ell$ signal regions shown in Figure~\ref{fig:cms-summ-allhadro}.
The measured cross section is $36\pm7\text{(stat)}^{+10}_{-8}\text{(syst)}$ fb and evidence for four top production is obtained
with an observed statistical significance over background of $3.9$ standard deviations.
The result is combined with an older CMS measurement in the $2\ell$SS, $3\ell$ and $4\ell$ final states~\cite{TOP-18-003},
leading to a measured cross section of $17\pm4\text{(stat)}\pm3\text{(syst)}$ fb.

\section{\texorpdfstring{The $2\ell$SS, $3\ell$ and $4\ell$ final states}{The 2lSS, 3l and 4l final states}}
Four top quark production was observed in the $2\ell$SS, $3\ell$ and $4\ell$ final states
by the ATLAS and CMS Collaborations~\cite{ATLAS_FourTop,TOP-22-013}.
The results are a reanalysis of the Run 2 datasets in these final states 
and supersede the previous results published by both collaborations~\cite{ATLAS_fourtop_SSDLML_old,TOP-18-003}.
The observations are made possible thanks to many improvements in reconstruction techniques, object identification such as b tagging, and analysis methods.

Compared to the previously discussed final states, the branching fraction is significantly smaller, at only 13\%,
but the background is drastically reduced as well and made up of a mix of reducible and irreducible processes.
One of the most important backgrounds are events, mainly \ttbar, with additional nonprompt leptons, 
which are leptons originating from semileptonic b decays or photon conversions in the detector material.
The rate of these events is in both results reduced by employing a per-lepton BDT, trained to identify prompt leptons.
A much smaller irreducible background comes from events with charge misidentified electrons.
In the ATLAS result, a per-electron BDT is also applied to reduce the misidentification rate, while in the CMS result,
three charge measurements are required to give the same result for each electron.
The majority of the reducible background is made up of \ttW{}, \ttZ{} and \ttbar H production.

\paragraph{ATLAS result}
The nonprompt background is split in 4 different contributions, predicted from MC simulations and normalized in four dedicated control regions.
In addition, a lot of care is taken on the \ttW{} background.
The main contribution is predicted with next-to-leading order (NLO) QCD MC and additive weights to consider the fully electroweak contributions.
Since it is known that the NLO term first order in QCD and third order in electroweak has a large contribution to the cross section,
a dedicated sample is simulated and used.
Lastly, the number of jets distribution is corrected using jet scaling regimes, where a ratio is defined based on the number of events
in neighbouring bins in the number of jets distribution. 
This ratio is expected to be equal to a constant at a large number of additional jets on top of the hard process,
and proportional to $1/(1+n)$, with $n$ the number of additional jets, for low values of n.
In practice, these two regimes flow over into one another.
Four control regions are defined to extract the parameters associated with these jet scaling regimes, 
as well as two normalization factors, one for \ttW$^+$ and one for \ttW$^-$ events.
The result of the corrections to the \ttW{} modelling are compared to data
using a number of jets distribution using bins of the number of events with two positive leptons minus the number
of events with two negative leptons, for which the result is shown in the left panel of Figure~\ref{fig:atlas-obs-res}.

\begin{center}
\includegraphics[width=0.44\textwidth]{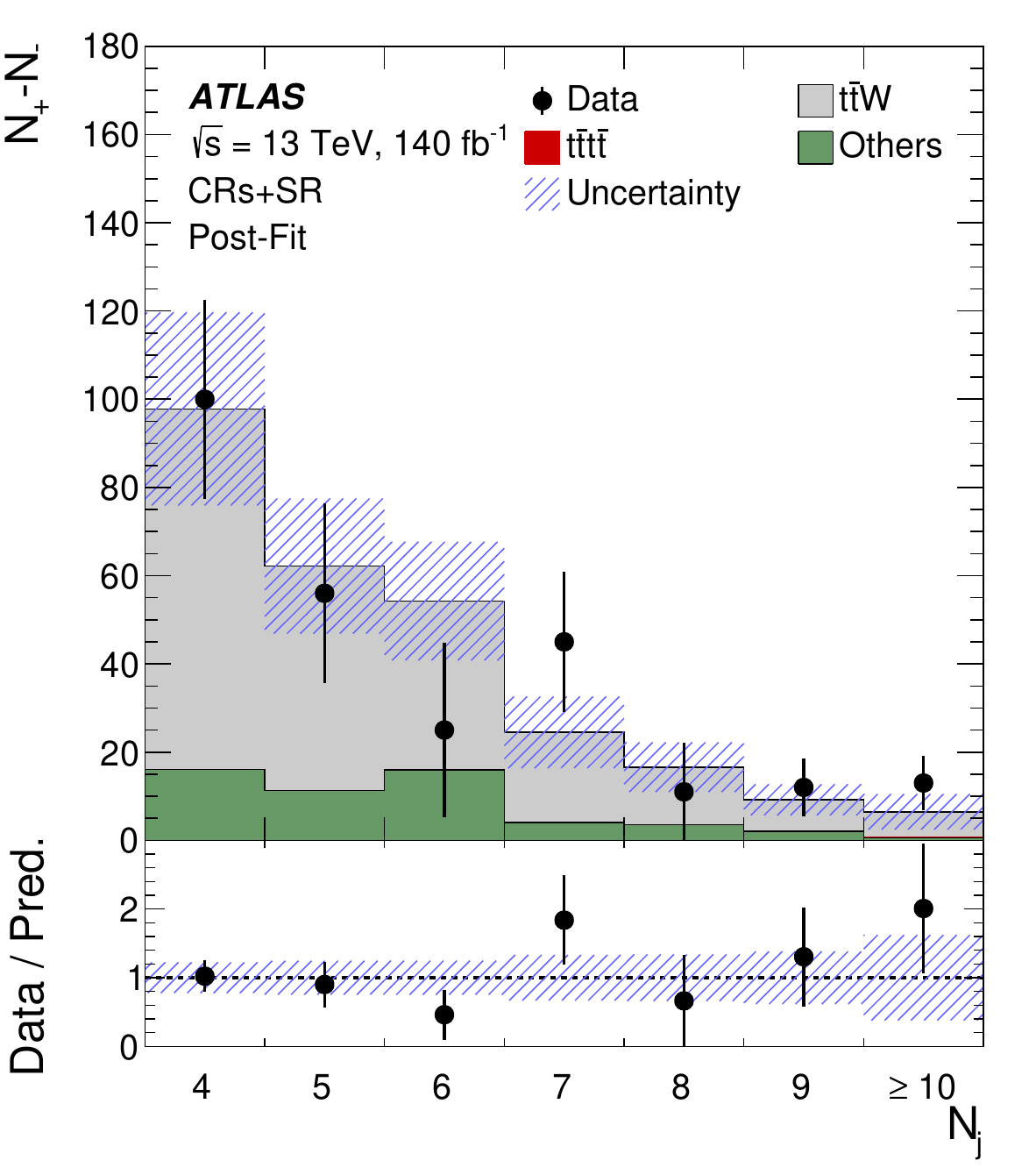}%
\hspace{0.06\textwidth}
\includegraphics[width=0.44\textwidth]{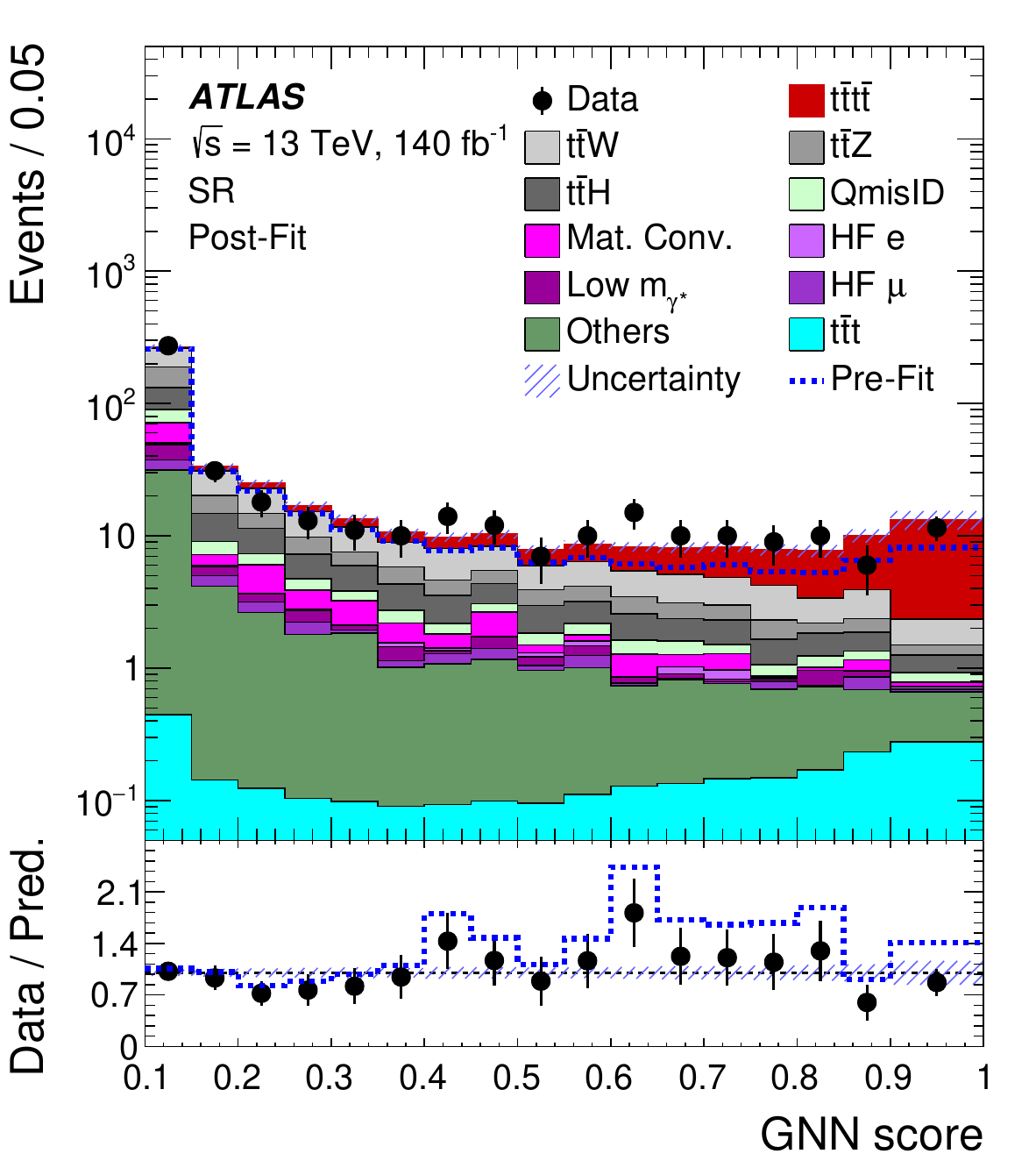}%
\captionof{figure}{%
    (left) Using the charge asymmetry in \ttW{} events, the modelling of the process as a function of the number of jets is checked 
    after the corrections with the jet scaling regimes~\cite{ATLAS_FourTop}.
    (right) The GNN output score in the signal region, used to extract the cross section of four top production~\cite{ATLAS_FourTop}.
}
\label{fig:atlas-obs-res}
\end{center}

In summary, a total of eight control regions are introduced.
To extract four top production, a single signal region is used, covering all considered final states, with a high requirement on
the number of jets, b tagged jets and the scalar sum of the transverse momenta of the jets and leptons in an event.
The signal is separated from the background processes by using a graph neural network (GNN) and the signal strength is extracted with
a template fit to the GNN score distribution in the signal region, as shown in Figure~\ref{fig:atlas-obs-res}, 
as well as one template per control region.
This leads to a measured cross section of $22.5^{+4.7}_{-4.3}\text{(stat)}^{+4.6}_{-3.4}\text{(syst)}$ fb 
and an observation of the process with a statistical significance of 6.1 standard deviations.

\paragraph{CMS result}
The nonprompt background is predicted using a tight-to-loose ratio method, where the contribution in the signal and control regions
is predicted from sidebands in data, defined by loose-but-not-tight leptons (where the signal and control regions are defined using tight leptons).

The \ttW{} background is predicted using NLO QCD MC with the normalization free floating in the fit. 
Two dedicated control regions in the same-sign dilepton final state are introduced, as well as using a the output of a multiclass BDT in the signal region,
which is trained to separate \ttW{} production from other background processes.

Dedicated signal regions are used for each of the analyzed four top final states.
Since the background composition changes between these final states,
three control regions are introduced for $3\ell$ and $4\ell$ events to ensure a correct modelling of the background.
Two of these regions are introduced for the $3\ell$ final state, one with a looser requirement on the number of (b tagged) jets,
and one requiring an opposite-sign same-flavour lepton pair with an invariant mass close to that of the Z boson.
In the $4\ell$ final state, a similar control region to the last one is used.
These two control regions allow to float the normalization of the \ttZ{} background in the fit, 
and also allow to check the contributions of WZ and ZZ production with additional (b) jets.

\begin{center}
\includegraphics[width=0.47\textwidth]{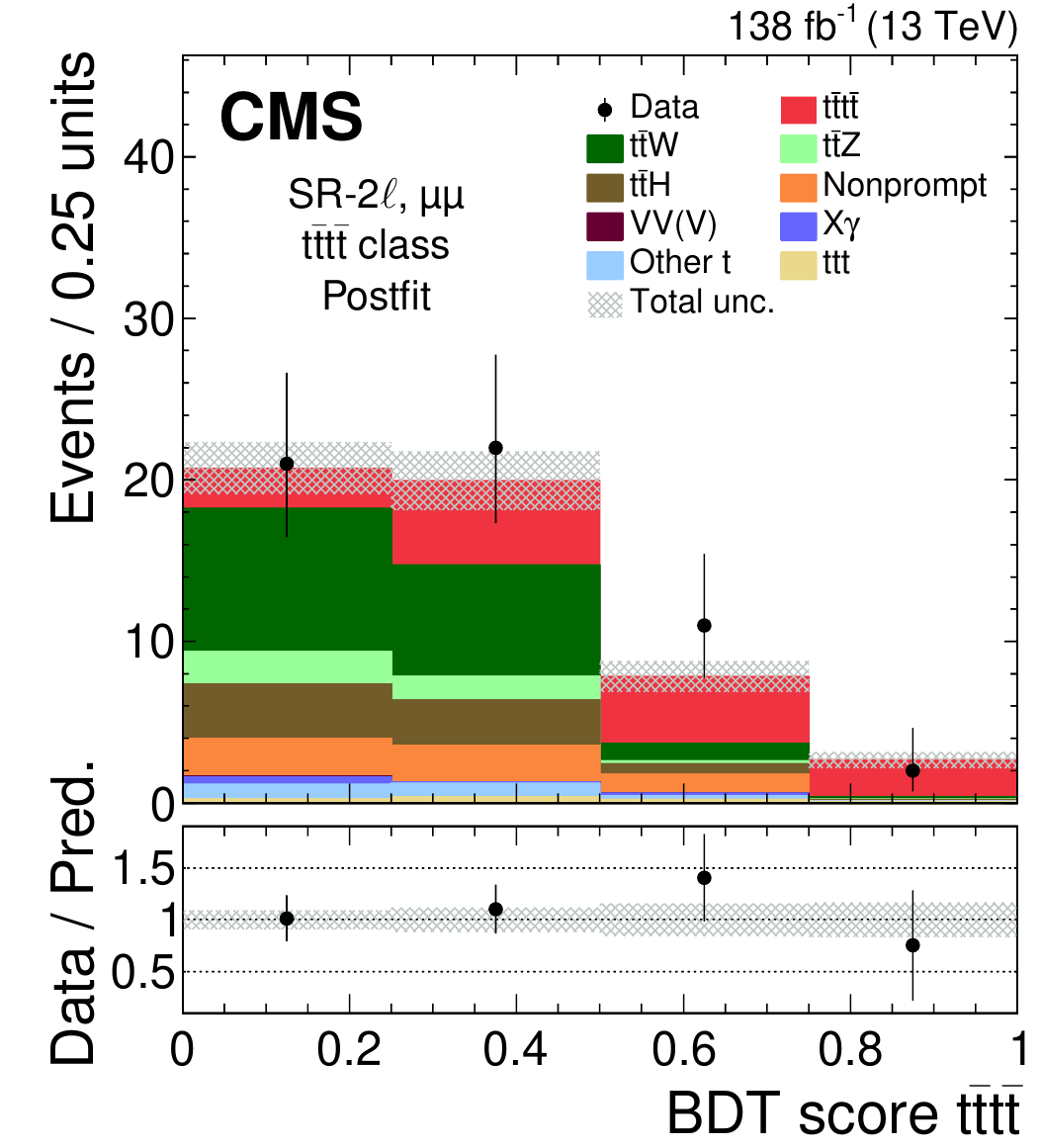}%
\hfill
\includegraphics[width=0.47\textwidth]{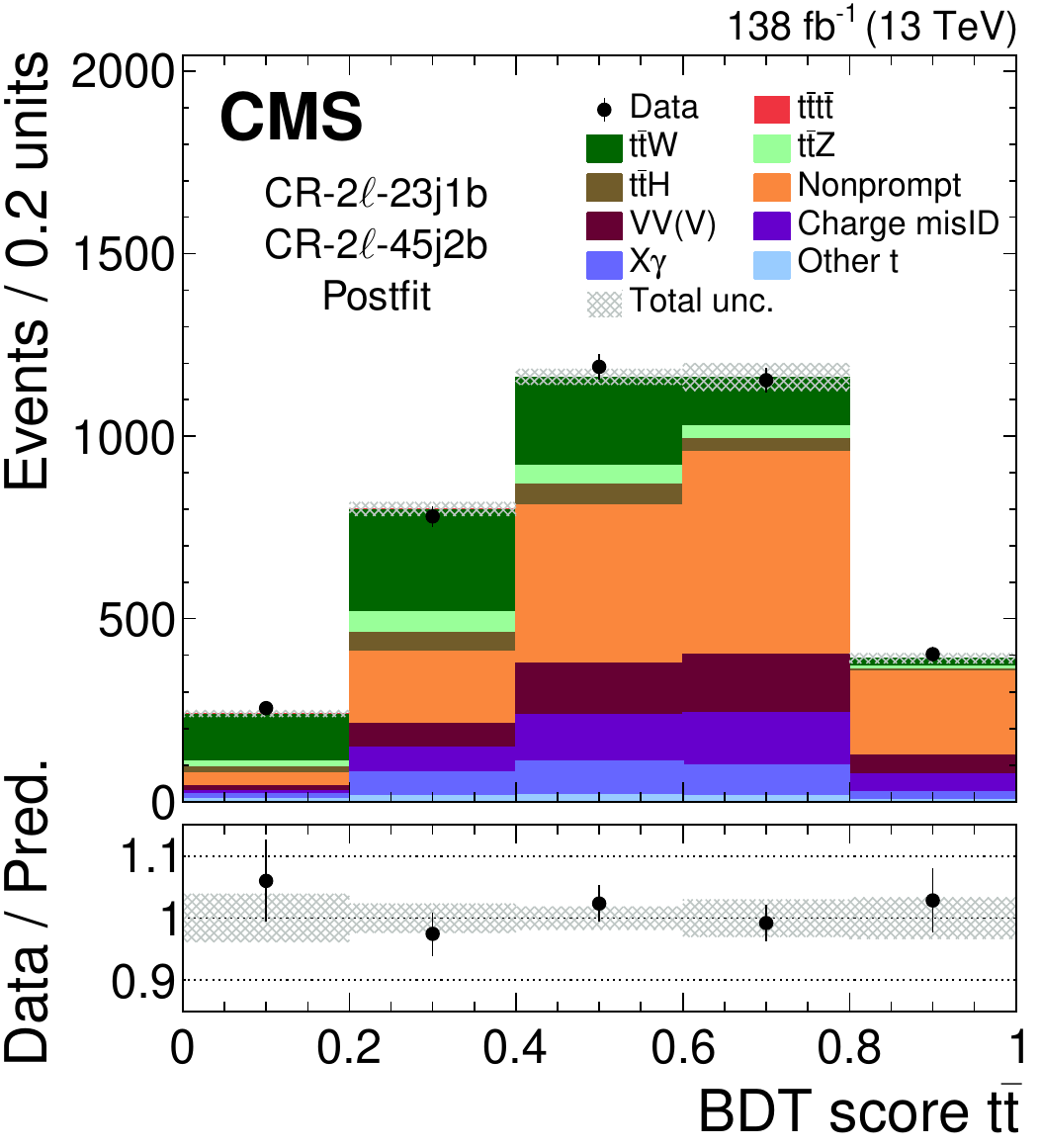}%
\captionof{figure}{%
(Left) An example of a signal region used in the fit to extract the four top cross section. The BDT score of the four top class is used to maximize
the sensitivity~\cite{TOP-22-013}.
(Right) Combination of the two same-sign dilepton control regions used in the analysis, designed to extract the \ttW{} cross section and
verify the nonprompt modelling. The signal region BDT is used to separate the \ttW{} and nonprompt contributions~\cite{TOP-22-013}.
}
\label{fig:cms-obs-res}
\end{center}
The signal regions are defined by a high number of (b tagged) jets and the scalar sum of the transverse momenta of all jets.
Two multiclass BDTs are trained, one for the same-sign dilepton and one for a combination of the three and four lepton final states.
Three classes are defined for each BDT: a signal class, a \ttbar{} class, trained with a combination of events with nonprompt and charge misidentified leptons, 
and a \ttbar X class, trained on a combination of \ttW, \ttZ{} and \ttbar H.
The signal regions are split in these three classes according to the BDT output and the same-sign dilepton signal region
is further split in electron-electron, electron-muon and muon-muon events.
The cross section is extracted with a template fit to a combination of signal and control regions,
with an example of these fitting regions for the signal and control regions shown in Figure~\ref{fig:cms-obs-res}.
The measured cross section is $17.7^{+3.7}_{-3.5}\text{(stat)}^{+2.3}_{-1.9}\text{(syst)}$ fb,
and the process is observed with a statistical significance of 5.6 standard deviations.

\section{Summary}
The most recent searches and measurements for four top quark production performed in various final states 
by the ATLAS and CMS Collaborations are presented. 
The process has been observed for the first time at both experiments but the statistical uncertainty
is still the limiting component.
The results all report a higher than expected cross section, but still agree 
with the standard model of particle physics within less than 1.8 standard deviations.

\end{document}